\documentclass[12pt]{article}
\usepackage{epsfig}
\usepackage{amsmath}
\usepackage{float}
\usepackage{subfigure}
\usepackage{a4}
\usepackage{float}

\newcommand{\eqs}[1]{\begin{eqnarray}#1\end{eqnarray} }
\newcommand{\ce}[1]{Eq.~(\ref{#1})}

\newcommand{\cf}[1]{{Figure~\ref{#1}}}

\newcommand{\ep}{\varepsilon}

\newcommand{\vect}[1]{\vec{#1}}

\newcommand{\ie}{{\it i.~e.}}
\newcommand{\eg}{{\it e.g.}}
\newcommand{\nn}{\nonumber}
\renewcommand{\cal}{\mathcal}

\begin{document}

\title{New contributions to \\ heavy-quarkonium production}

\author{\vspace*{0.2cm}J.P. Lansberg$^{a,b}$, J.R. Cudell$^a$ and Yu. L. Kalinovsky$^{a,c}$\\
\centerline{\small $^a$ \it Physique th\'eorique fondamentale, D\'ep. de  Physique, 
Univ. de  Li\`ege, }\\
\centerline{\small \it ~~All\'ee du 6 Ao\^{u}t 17, b\^{a}t. B5a, B-4000 Li\`ege~1, Belgium}\\
\centerline{\small $^b$ \it Centre de Physique Th\'eorique, \'Ecole Polytechnique\protect\footnote{Unit\'e mixte 7644 du CNRS},}\\
\centerline{\small \it ~~ F-91128 Palaiseau, France}\\
\centerline{\small $^c$ \it  Lab. of Information Technologies, Joint Inst. for Nuclear Research, }\\
\centerline{\small \it ~~Dubna, Russia}\\
\centerline{\small \it E-mails: Jean-Philippe.Lansberg@cpht.polytechnique.fr, JR.Cudell@ulg.ac.be,}\\
\centerline{\small \it kalinov@qcd.theo.phys.ulg.ac.be}
}
\date{}
\maketitle

\begin{abstract}
We reconsider quarkonium production in a field-theoretical setting and 
we show that the lowest-order mechanism for
heavy-quarkonium production receives in general contributions from two 
different cuts. The first one corresponds to the usual 
colour-singlet mechanism. 
The second one has not been considered so far. We treat it in a gauge-invariant 
manner, and introduce new 4-point vertices, suggestive of the colour-octet
mechanism. These new objects enable us to go beyond the static approximation.
We show that the contribution of the new cut can be as large as the usual
colour-singlet mechanism at high $P_T$ for $J/\psi$. In the $\psi'$ case,
theoretical uncertainties are shown to be large and agreement with data is
possible.
\end{abstract}

\newpage
\section{Introduction}
Years after the first disagreement between data~\cite{CDF7997a,CDF7997b} 
and the colour-singlet model (CSM)~\cite{CSM_hadron,CSM_frag}, the problem of 
heavy-quarkonium production --in particular
at the Tevatron-- is still with us. Indeed, it is widely accepted now that fragmentation 
processes, through the colour-octet 
mechanism (COM)~\cite{COM} 
dominate the production of heavy quarkonia in high-energy hadronic collisions, 
even for $P_T$ as low as 6 GeV. 
In the COM, one then 
parametrises the non-perturbative transition from octet to singlet by unknown
matrix elements, which are determined to reproduce the data. However, fragmentation
processes are known to produce mostly transversely-polarised vector mesons 
at large transverse momentum~\cite{Cho:1994ih} 
and, in the $J/\psi$ and $\psi'$ cases, this seems in contradiction with the 
 measurements from CDF~\cite{Affolder:2000nn}. For a comprehensive review
on the subject, the reader may refer to~\cite{yr_QWG}.

We therefore reconsider the basis of quarkonium ($\cal Q$) production in field
theory, and concentrate on $J/\psi$ and $\psi'$ production. 
We shall see that new contributions are present in the
lowest-order diagrams, and we shall also explain how one can build
a consistent and systematic scheme to go beyond the static approximation. 
To this end, we shall use 3-point vertices depending on the relative momentum 
of the constituent quarks and normalised to the leptonic width of the meson. 
We shall show that, in order to preserve gauge invariance, it is 
required to introduce vertices more complicated than the 
3-point vertex.

Finally, we shall see that our formalism can be easily applied to the production
of excited states. In the case of $\psi'$, the theoretical uncertainties 
are unexpectedly large and allow agreement with the data. 

\section{Bound states in QCD}
All the information
needed to study processes involving bound states, 
such as decay and production mechanisms, can be parametrised by vertex functions, which describe
the coupling of the bound state to its constituents and contain the information 
about its size, the amplitude of probability for given quark configurations 
and the normalisation of its wave function. In the case of heavy quarkonia, 
the situation simplifies as they can be approximated by their lowest Fock state, made of 
a heavy quark and an antiquark, combined to obtain the proper quantum numbers. 
Furthermore, it has been shown~\cite{Burden:1996nh} that, for light {\sl vector} 
mesons, the dominant projection operator is $\gamma^\mu$ and we expect this to hold even better
for heavy vector mesons as this approximation gets better in the case of 
$\phi(s\bar s)$.

The transition 
$q\bar q\rightarrow {\mathcal Q}$ can then be described by the following
 3-point function:
\begin{eqnarray}\label{vf}
\Gamma^{(3)}_{\mu}(p,P) = \Gamma(p,P) \gamma_\mu,
\end{eqnarray}
with $P\equiv p_{1}-p_{2}$ the total momentum of the bound state, and $p\equiv(p_{1}+p_{2})/2$ 
the relative momentum of the bound quarks, as drawn in \cf{fig:BS_vertex_phenoa}. This choice
amounts to describing the vector meson as a massive photon with a non-local
coupling. 

\begin{figure}[t]
\centering
\includegraphics[width=10cm]{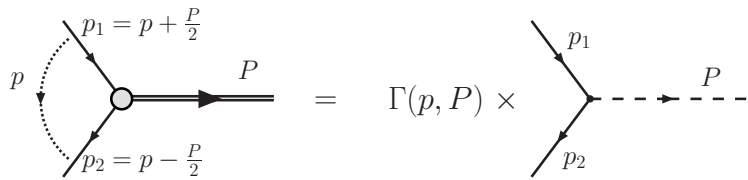}
\caption{Bound-state vertex obtained by multiplying a point vertex, representing a structureless
particle, by a form factor.}
\label{fig:BS_vertex_phenoa}
\end{figure} 

We do not assume that the quarks
are on-shell : their momentum distribution comes from $\Gamma(p,P)$ and
from their propagators. In order to make contact with wave functions,
and also to simplify calculations, we assume that $\Gamma(p,P)$ can
be taken as a function of the square of the
relative c.m. 3-momentum $\vec p$ of the quarks, which can be written in a 
Lorentz invariant form as $\vec p^{\,\, 2}=
-p^2+\frac{(p.P)^2}{M^2}$. For the functional form of $\Gamma(p,P)$, we neglect
possible cuts, and choose two otherwise extreme scenarios:
a dipolar form which decreases slowly with $\vect p$, and a Gaussian form: 
\eqs{
\Gamma(p,P)=\frac{N}{(1+\frac{\vec p^{\,2}}{\Lambda^2})^2} \text{ and } \Gamma(p,P)=N e^{-\frac{\vec p^{\,2}}{\Lambda^2}} 
,}
both with a normalisation $N$ and a size parameter $\Lambda$, which
can be obtained from relativistic quark models~\cite{Lambda}.
We shall see in Section 4 how we fix the latter using the 
leptonic-decay width.

\section{Lowest-order production diagrams}
In high-energy hadronic collisions, quarkonia are produced at small $x$,
where protons are mainly made of gluons. Hence
gluon fusion is the main production mechanism. 
In the case of $J/\psi$ and $\psi'$,
a final-state gluon emission is required to conserve $C$ parity. This gluon also
provides the $\cal Q$ with transverse momentum $P_T$. We assume here that
we can use collinear factorisation to describe the initial gluons, and hence
that the final-state gluon emission is the unique source of $P_T$. All the relevant 
diagrams for the lowest-order gluon-initiated production process 
can be obtained from that of \cf{fig:Landau} by crossing. There are six of them.

\begin{figure}[t]
\centerline{\includegraphics[width=4cm]{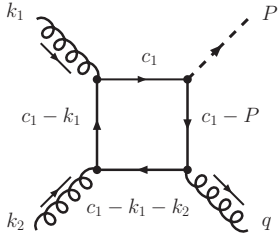}}
\caption{Box diagram.}
\label{fig:Landau}
\end{figure}

These diagrams have discontinuities, which generate their imaginary parts.
In order to find these, we can use the Landau equations~\cite{Landau_eq}.
It is sufficient to consider the diagram of \cf{fig:Landau}, 
for which the Landau equations become:
\begin{eqnarray}
\lambda_1 (c_1^2-m^2)&=&0\nonumber\\
\lambda_2 ((c_1-P)^2-m^2)&=&0\nonumber\\
\lambda_3 ((c_1-k_1)^2-m^2)&=&0\\
\lambda_4 ((c_1-k_1-k_2)^2-m^2)&=&0\nonumber\\
\lambda_1 c_1+\lambda_2 (c_1-P)+\lambda_3 (c_1-k_1)+\lambda_4 (c_1-k_1-k_2)&=&0\nonumber
\label{eq:landau}
\end{eqnarray}
with $m$ the quark mass.
These equations have only two solutions in the physical region: 
one which is always present and gives a cut which starts at $\hat s=(k_1+k_2)^2=4 m^2$, 
shown in \cf{fig:diag_LO_QCD}~(a), corresponding to a cut through the two $s$-channel 
propagators, and another one when the meson mass $M$ is larger or 
equal to $2 m$ (see \cf{fig:diag_LO_QCD}~(b): this corresponds to a cut through 
the two propagators touching the meson). The latter leads to the colour-singlet 
model~\cite{CSM_hadron}, which assumes that the quarks should be put on-shell 
to make the meson. The former cut has not been considered so far for the 
description of inclusive production. Let us mention however that
similar cuts are dominant in diffractive production
of vector mesons~\cite{diffractive}, or in DVCS~\cite{DVCS}.

We are going to consider this $s$-channel cut in detail.  To avoid 
complications, we choose quark masses $m>M/2$ high enough for the second cut  
not to contribute. This will also simplify the normalisation procedure 
for the vertex.

\begin{figure}[H]
\centerline{\mbox{
\subfigure[cut 1]{\includegraphics[height=3.5cm]{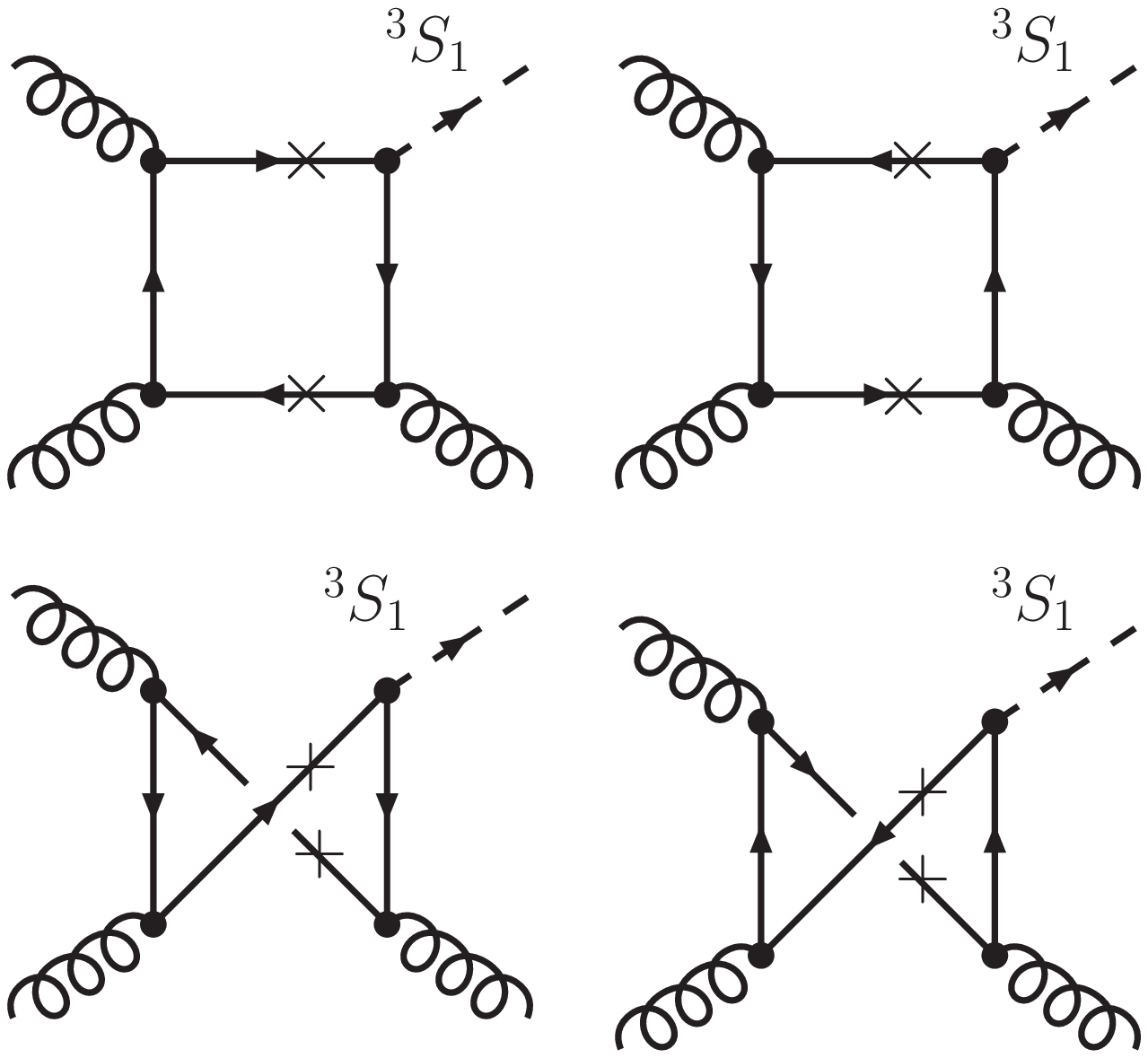}}
\quad\quad\quad
\subfigure[cut 2]{\includegraphics[height=3.5cm]{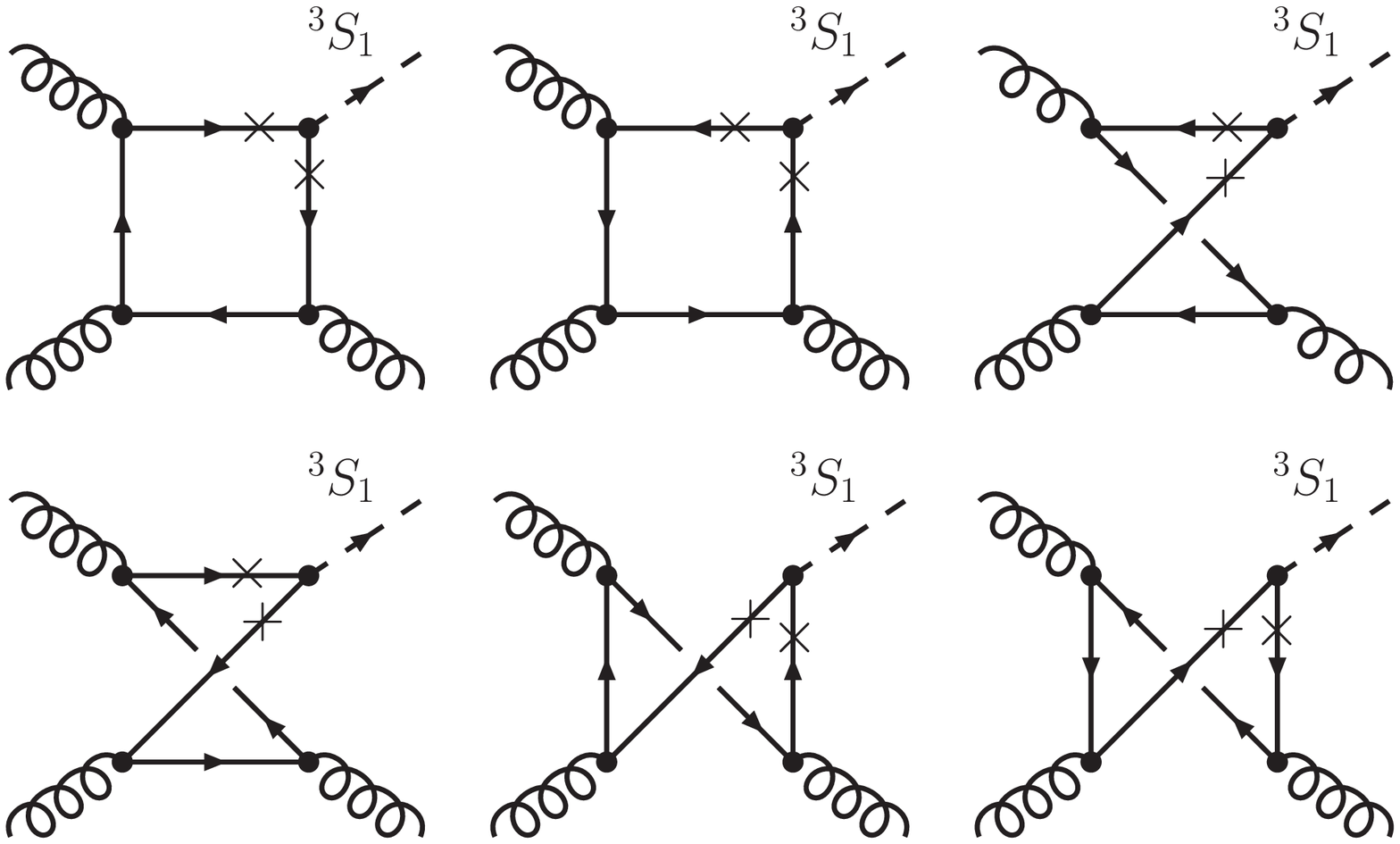}}
}}
\caption{The first family (a) has 4 diagrams 
and  the second family (b) 6 diagrams 
contributing the discontinuity of $gg \to \!\!\ ^3S_1 g$ at LO in QCD.}
\label{fig:diag_LO_QCD}
\end{figure} 

\subsection{Non-locality and gauge invariance}
The first problem one immediately faces when evaluating the diagrams of\break
\cf{fig:diag_LO_QCD}~(a) is that of gauge invariance: whereas these
diagrams are gauge invariant if we have a photon instead of a $\cal Q$, 
they are not for a finite-size object. Indeed, the vertex function 
$\Gamma(p,P)$ takes different values in diagrams where either the on-shell 
quark or the antiquark  touches the $\cal Q$: the relative momentum 
$p$ is then either $p=2c_1-P$ or $p=2c_2+P$, so that the 
delicate cancellation that ensures current conservation is spoilt.

\begin{figure}[H]
\centerline{\mbox{\subfigure[]{\includegraphics[height=4.25cm]{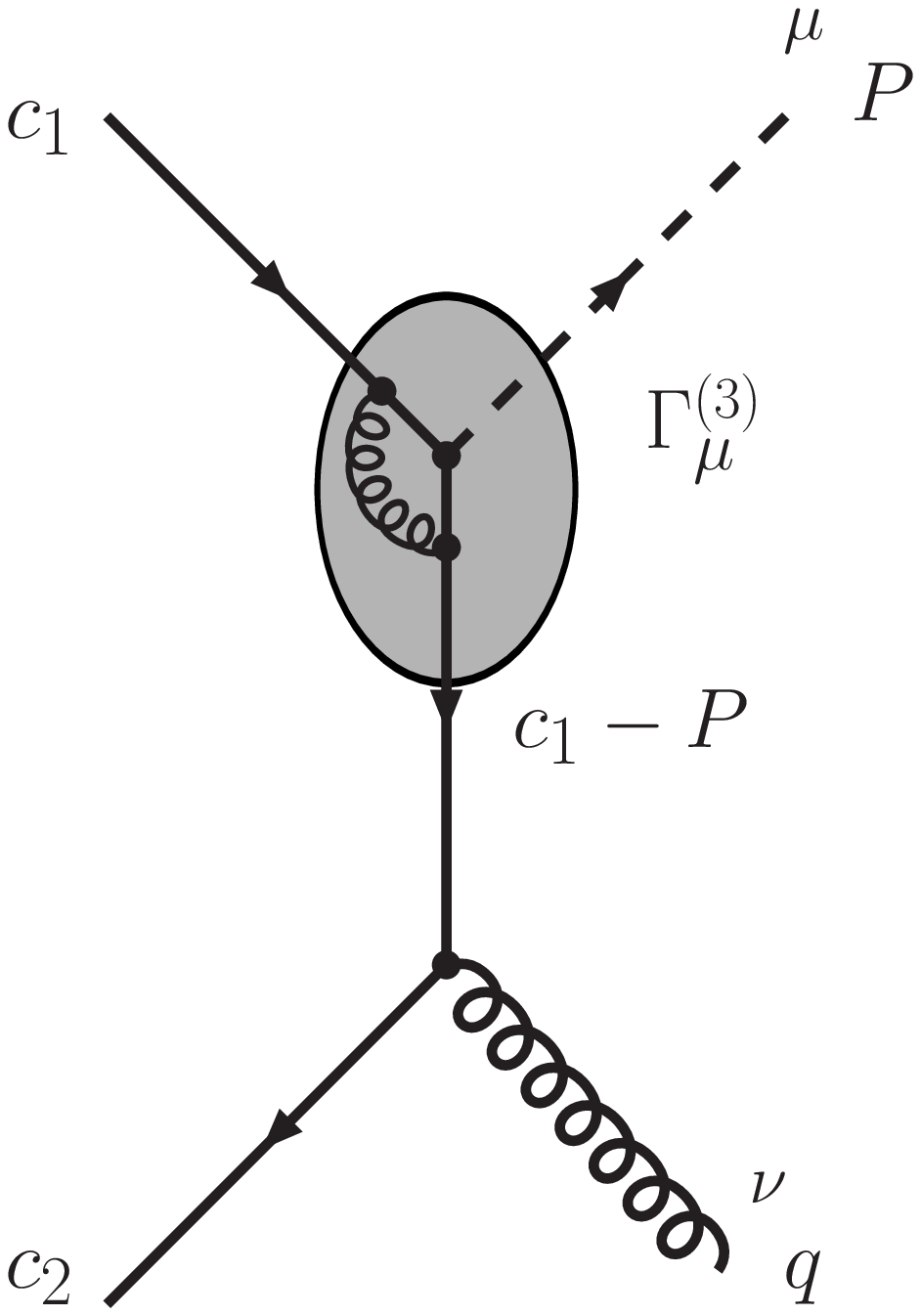}\quad\quad
\includegraphics[height=4.25cm]{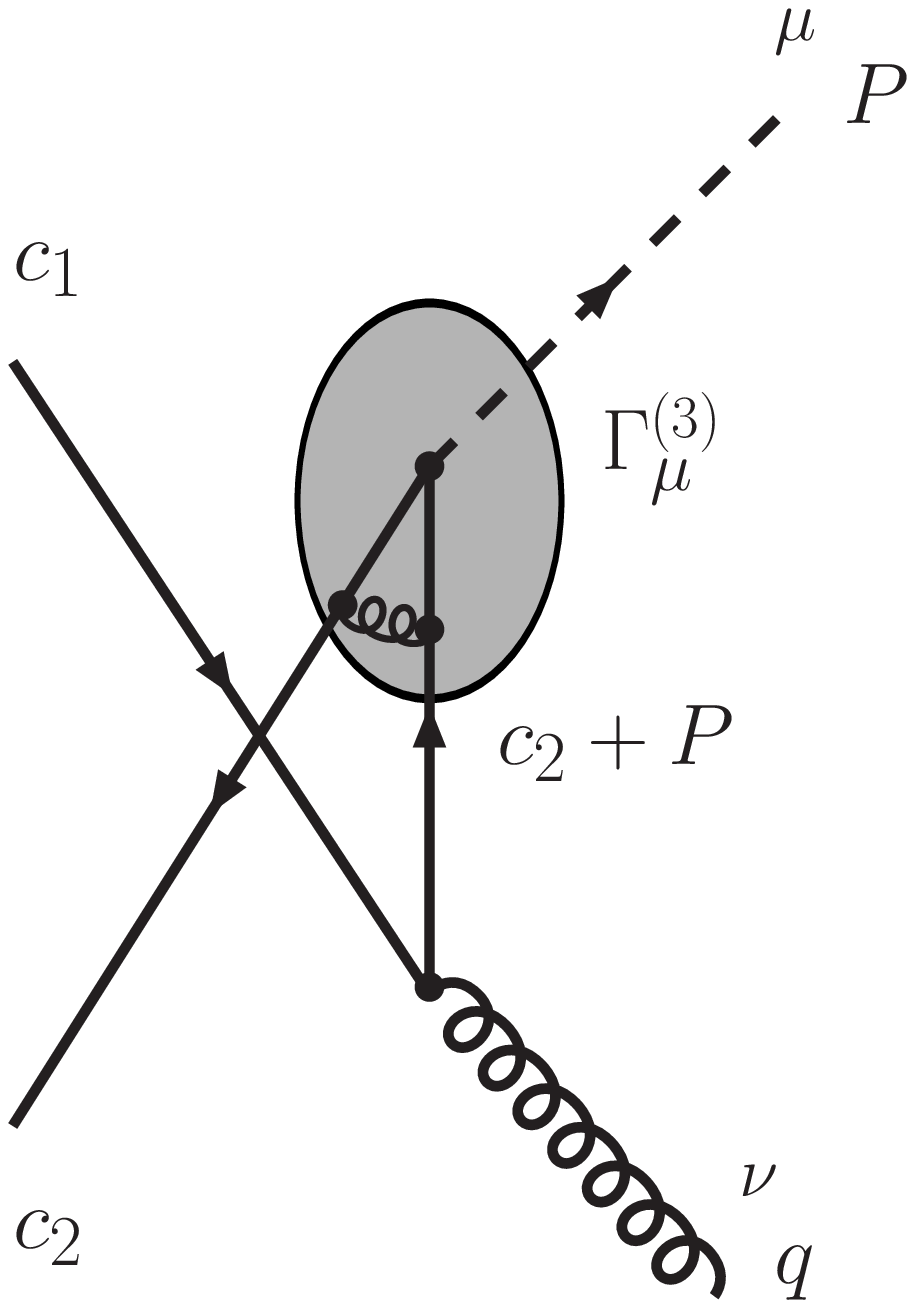}}\quad\quad
\subfigure[]{\includegraphics[height=4.25cm]{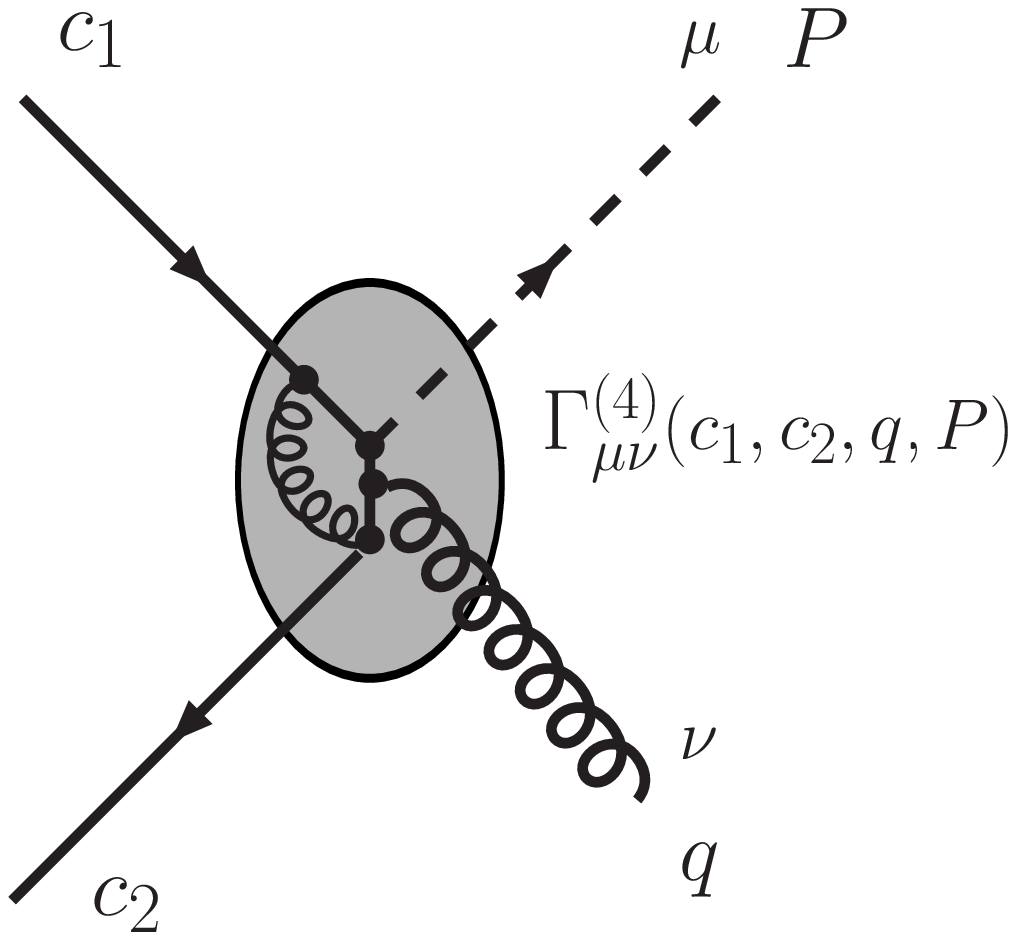}}}}
\caption{Illustration of the necessity of a 4-point vertex.}
\label{fig:illus_GI_break}
\end{figure} 

The reason is easily understood:
if one considers a local vertex, then the gluon can only couple to the
quarks that enter the vertex. For a non-local vertex, it
is possible for the gluon to connect to the quark or gluon lines inside the vertex,
as shown in \cf{fig:illus_GI_break}\footnote{In the cases of~\cf{fig:diag_LO_QCD} (b), 
for $M=2m$, no gluon emission is kinematically allowed and the diagrams are directly
gauge-invariant.}.
These contributions must generate a 4-point $q\bar q {\cal Q}g$ 
vertex, $\Gamma^{(4)}_{\mu\nu}(c_1,c_2,q,P)$.  In
general, its form is unknown, but it must obey some general constraints ~\cite{Drell:1971vx,Lansberg:2005aw}:
\begin{itemize}
\item
it must restore gauge invariance: its addition to the amplitude must
lead to current conservation at the gluon vertex;
\item
it must obey crossing symmetry (or invariance by $C$ conjugation) which 
can be written
\begin{equation}
\Gamma^{(4)}_{\mu\nu}(c_1,c_2,q,P,m)=-\gamma_0\Gamma_{\mu\nu}^{(4)}(-c_2,-c_1,q,P,-m)^\dagger\gamma_0;
\end{equation}
\item
it must not introduce new singularities absent from the propagators or 
from $\Gamma(p,P)$, hence it can only have denominators proportional to $(c_1-P)^2-m^2$ or
$(c_2+P)^2-m^2$;
\item
it must vanish in the case of a local vertex $\Gamma_\mu^{(3)}\propto\gamma_\mu$, 
hence we multiply it by $\Gamma(2c_1-P,P)-\Gamma(2c_2+P,P)$.
\end{itemize}

These conditions are all fulfilled by the following choice~\cite{Lansberg:2005aw}:
\begin{eqnarray}
\Gamma^{(4)}_{\mu\nu}(c_1,c_2,P,q)&=&-i g_s T^a_{ki} \left[ \Gamma(2c_1+P,P)
-\Gamma(2c_2-P,P)\right]\nonumber \\
&\times&\left[\frac{c_{1\nu}}{(c_2+P)^2-m^2}
+\frac{c_{2\nu}}{(c_1-P)^2-m^2}\right]\gamma_\mu
\label{Gamma4}
\end{eqnarray}
where the indices of the colour matrix $T$
are defined in \cf{fig:gamma_4}, and $g_s$ is the strong coupling constant.

\begin{figure}[H]
\centering
\includegraphics[width=3cm]{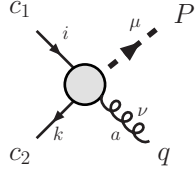}
\caption{The gauge-invariance restoring vertex, $\Gamma^{(4)}$.}
\label{fig:gamma_4}
\end{figure} 

It must be noted first that the quark pair $(c_1,c_2)$ that makes the meson is now in
a colour-octet state. We thus recover the necessity of such configurations, in this case to restore gauge invariance. 
We must also point out that this choice of vertex 
is not unique. We postpone a full study of the 4-point vertex ambiguity~\cite{Lansberg:2005aw} to another paper. 

When taken into account
in the calculation of the discontinuity of $gg\to {\cal Q} g$, 
$\Gamma^{(4)}$ introduces two new diagrams\footnote{The 
contributions of the triple-gluon vertex on the left and 
$\Gamma^{(4)}$ on the right is zero due to
charge conjugation.} shown in \cf{fig:LO_new_diag}. 
Including these contributions in the calculation
of the amplitude, we obtain a gauge-invariant quantity.

\begin{figure}[H]
\centering
\includegraphics[width=8cm]{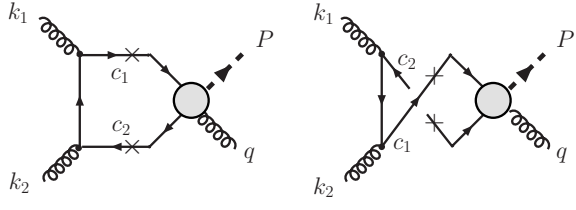}
\caption{New contributions to the discontinuity of the amplitude from $\Gamma^{(4)}$.}
\label{fig:LO_new_diag}
\end{figure} 

\section{Amplitudes}
We define ${\cal A}_i^{\mu\nu\rho\sigma}$, $i=1,$..., 6,
as the unintegrated amplitude
given by the usual Feynman rules for the four diagrams 
of~\cf{fig:diag_LO_QCD}~(a) and the two of~\cf{fig:LO_new_diag}. We choose
the loop momentum $\ell$ so that $c_1= \ell+k_1$ and $c_2=\ell-k_2$. We then have for
the imaginary part of the physical amplitude ${\cal M}$:
\eqs{
{\cal M}^{pqrs}&=& \frac{1}{2}  \sum_{i=1}^6 \int \frac{d^4\ell}{(2\pi)^4} {\cal A}_i^{\mu\nu\rho\sigma}
\ep^p_{1,\mu}\ep^q_{2,\nu} \ep^{r\star}_{3,\rho} \ep^{s\star}_{4,\sigma} \nn \\ 
&\times &2\pi\delta^{+}((\ell+k_{1})^{2}-m^{2} ) 2\pi \delta^{-}((\ell-k_{2})^{2}-m^{2})\nn \\
}

This polarised partonic amplitude is thus obtained by contracting 
${\cal A}_i^{\mu\nu\rho\sigma}$ with the 
polarisation vectors of the gluons $\ep^p_{1,\mu}$, $\ep^q_{2,\nu}$, $\ep^s_{4,\sigma}$ and with that of 
the vector meson $\ep^r_{3,\rho}$, by integrating on 
the internal phase space restricted by the cutting rules and 
by summing the six contributions from the
diagrams of \cf{fig:diag_LO_QCD}~(a) and \cf{fig:LO_new_diag}. 

The complete expressions of the polarisation vectors are as follows.
One of the transverse polarisations can be taken as orthogonal to the plane of the collision:
\begin{eqnarray}
\ep_{1}^{T_1}=\ep_{2}^{T_1}=\ep_{3}^{T_1}=\ep_{4}^{T_1}=\epsilon_T,
\end{eqnarray} with
$\epsilon_T.k_1=\epsilon_T.k_2=\epsilon_T.P=0$. The other transverse 
polarisation can be taken as
\begin{eqnarray}
\ep_{1}^{T_2}=\sqrt{\frac{1}{\hat s \hat t \hat u}}\left(\hat t k_1+ \hat u k_2+ \hat s q\right)=\ep_{2}^{T_2}
\end{eqnarray} for the two gluons, and as 
\eqs{\label{eq:pol-final2}
\ep_{3}^{T_2}=\sqrt{\frac{1}{\hat s \hat t \hat u}}\left(\hat t k_1-\hat u k_2+\left(\frac{\hat s(\hat t-\hat u)}{\hat t+\hat u}\right) q\right)
=\ep_{4}^{T_2},\\
} for the final-state gluon and for the vector meson, 
where $\hat s=(k_1+k_2)^2$, $\hat t=(k_2-q)^2$ and $\hat u=(k_1-q)^2$
are the Mandelstam variables for the partonic process.

Finally, the longitudinal
vector-meson polarisation can be taken as
\eqs{
\ep_{3}^{L}  = \frac{1}{M}\left(k_1+k_2-\left(\frac{\hat s+M^2}{\hat s-M^2}\right) q\right),\\
}

To complete the calculation of the amplitude, we need to normalise
the 3-point vertex function $\Gamma^{(3)}$. We use here the leptonic 
decay width to fix this normalisation~\cite{Lansberg:2005aw,Lansberg:2005ed}.

The width in terms of the decay amplitude is given by
\eqs{\label{eq:lept_width}
\Gamma_{\ell\ell}=\frac{1}{2 M}\frac{1}{(4\pi^2)}\int \left|\bar{\cal{M}}\right|^2 d_2(PS),
}
where $d_2(PS)$ is the two-particle phase space~\cite{barger}.

The amplitude is obtained as usual through Feynman rules. At lowest order in $\alpha_s$,
only the 3-point vertex function needs to be considered (giving an explicitly gauge invariant
answer). The square of the amplitude is then obtained from the diagram drawn in~\cf{fig:decay_diag_1}.

\begin{figure}[H]
\centering{\mbox{\includegraphics[width=10cm]{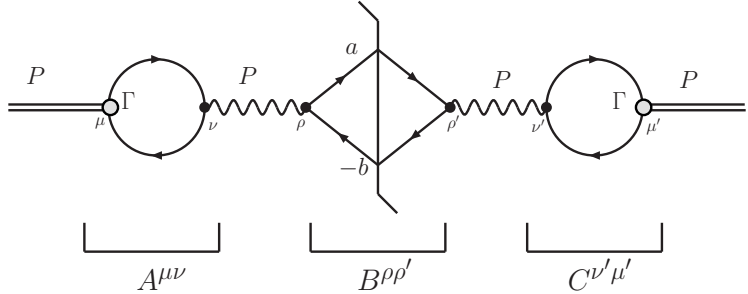}}}
\caption{Feynman diagram for $^3S_1\to \ell \bar \ell$.}\label{fig:decay_diag_1}
\end{figure}

\noindent In terms of the sub-amplitudes $A^{\mu\nu}$, $B^{\mu\nu}$ and $C^{\mu\nu}$ defined in
\cf{fig:decay_diag_1}, we have\footnote{We performed the calculation in the Feynman gauge, but the results are gauge invariant.}:
\eqs{
\label{eq:decomp_ampl_inv_decay}
\int \left|\bar{\cal M}\right|^2 d_2(PS)=-\frac{1}{3} 
\Delta_{\mu\mu'}
A^{\mu\nu}\frac{g_{\nu\rho}}{M^2}B^{\rho\rho'}\frac{g_{\rho'\nu'}}{M^2}
C^{\nu'\mu'},
}
where the factor $\Delta_{\mu\nu}=(g_{\mu\nu}-\frac{P_\mu P_{\nu'}}{M^2})=
\sum_i \ep_{i,\mu} \ep^\star_{i,\mu'}$ comes from the sum over polarisations 
of the meson and the factor $\frac{1}{3}$  from the averaging over the initial polarisations.

$B^{\rho\rho'}$, after 
integration on the two-particle phase space,  is found to be
\eqs{
B^{\rho\rho'}=-e^2 \frac{2\pi}{3} M^2\underbrace{\left[g^{\rho\rho'}- 
\frac{P^\rho P^{\rho'}}{M^2}\right]}_{\Delta^{\rho\rho'}}.
}
$A^{\mu\nu}$  (or equivalently ${C^{\mu\nu}}^\dagger$) can be written
(see \cf{fig:decay_diag_1a}):
\eqs{\label{eq:Amunu_start}
iA^{\mu\nu}=-3e_Q\int\! \frac{d^4k}{(2 \pi)^4}
 \Gamma(k,P)\frac{g^{\mu\nu} (M^2+4m^2-4k^2)+8 k^\mu k^\nu-2P^\mu P^\nu}
{((k-\frac{P}{2})^2-m^2+i\ep)((k+\frac{P}{2})^2-m^2+i\ep)},
}
with $e_Q$ the heavy-quark charge.

\begin{figure}[H]
\centering{\mbox{\includegraphics[width=5cm]{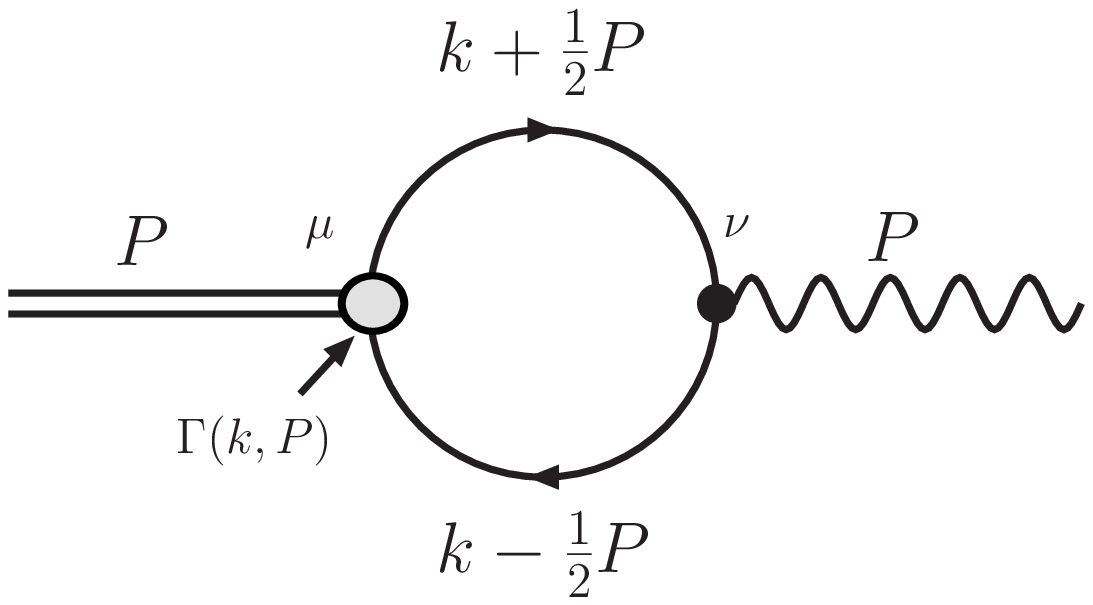}}}
\caption{Feynman diagram for $^3S_1\to \gamma^\star$.}\label{fig:decay_diag_1a}
\end{figure}
Performing the $k^0$ integration by residues, one obtains
\eqs{A^{\mu\nu}=\frac{- e_Q}{\pi^2} N I(\Lambda,M,m) \Delta^{\mu\nu}.
}
with
\eqs{\label{eq:I}
I(\Lambda,M,m)\equiv\int_0^\infty \!\! \frac{dK K^2\Gamma(k,P)}{\sqrt{K^2+m^2}N}
\frac{(2K^2+3m^2)}{(K^2+m^2-\frac{M^2}{4})}.
}
and $K=|\vec k|$.
$I$ is a function of $\Lambda$ through the vertex function $\Gamma(k,P)$ 
and is not in general computable analytically, but it is straightforward to get its
numerical value.

\begin{figure}[H]
\centering\includegraphics[width=0.49\textwidth]{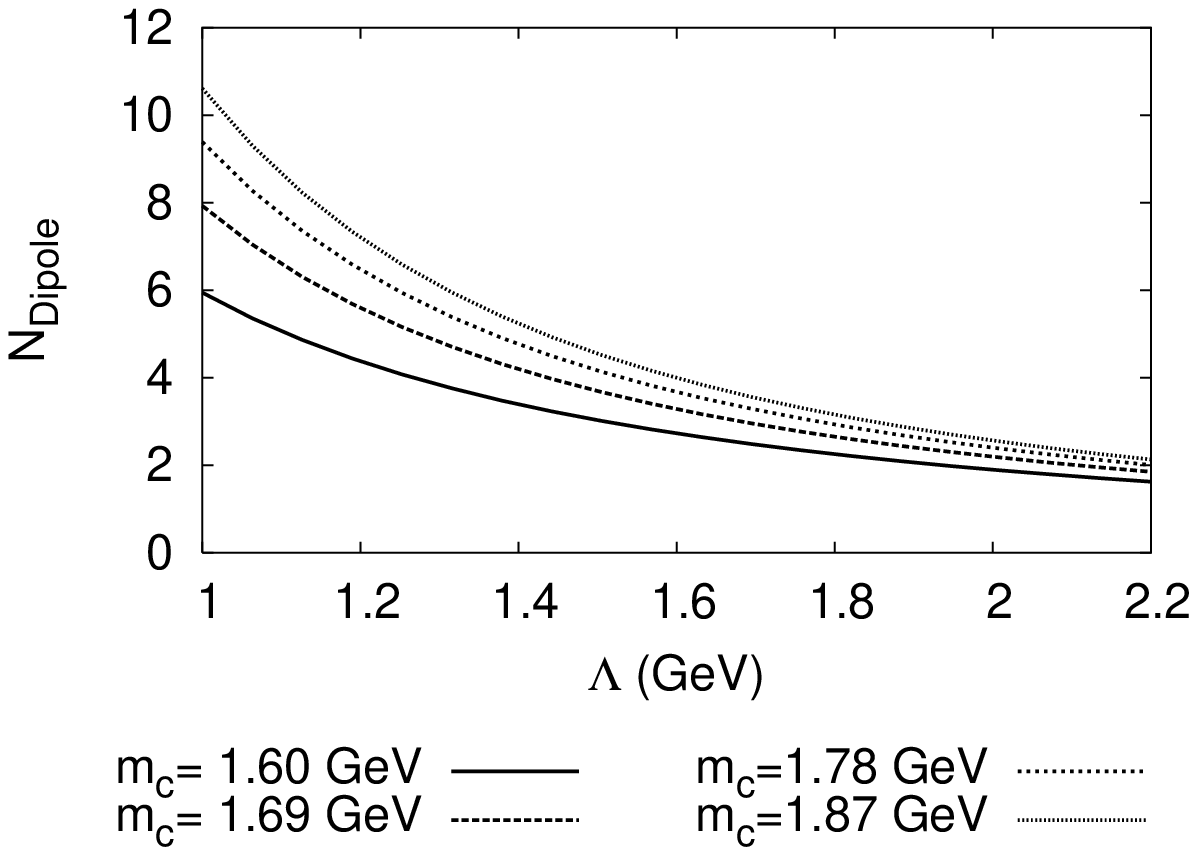}
\centering\includegraphics[width=0.49\textwidth]{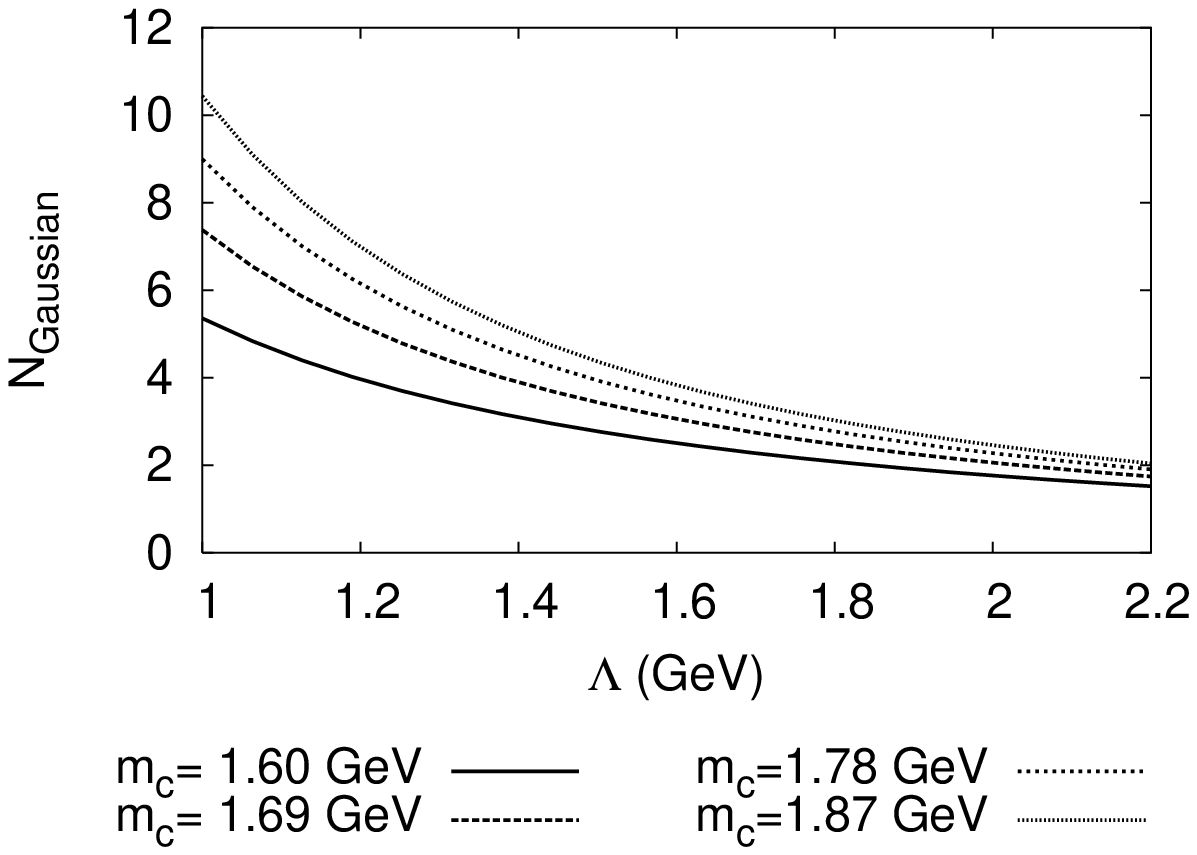}
\caption{Normalisation for a dipole (resp. Gaussian) form of $\Gamma(p,P)$ in the $J/\psi$ case 
as a function of $\Lambda$: left (resp. right).}
\label{fig:N_jpsi_lam}
\end{figure}
We can then put all the pieces together using Eqs. (\ref{eq:lept_width}) and (\ref{eq:decomp_ampl_inv_decay}) 
to determine $N$ from the measured leptonic width. 
We show in \cf{fig:N_jpsi_lam} the result in the $J/\psi$ case. 
As can be seen, the normalisation of the vertex depends rather strongly on $m_c$
and $\Lambda$, but very little on the assumed functional
dependence of the vertex function. We shall see later that once $N$ is
determined from the leptonic rate, the production cross section depends
little on these uncertainties.

\section{Production cross sections}

We can now evaluate the production cross section from the $s$-channel cut.
As stated before, we assume that collinear factorisation can be used,
in which case the link between the partonic and the 
hadronic cross sections is given by the following general formula:
\begin{equation}
\label{eq:Edsdp3}
E\frac{d^3\sigma}{dP^3}=\int_0^1 dx_1 dx_2 \ g(x_1) \ g(x_2)\ \frac{\hat s}{\pi}\frac{d\sigma}{d\hat t}
\delta(\hat s +\hat t +\hat u -M^2),
\end{equation}
where $x_1$ and $x_2$ are the momentum fractions of the incoming gluons, 
$P=(E,\vect P)$ is the momentum of the meson in the c.m. frame 
of the colliding hadrons, $g(x)$ is the gluon distribution function\footnote{In
 our calculations, we have chosen two  LO gluon parametrisations, 
MRST~\cite{Martin:2002dr} and CTEQ~\cite{Pumplin:2002vw}. For
each plot, the one used will be specified.}  taken
at the scale $\sqrt{M^2+P_T^2}$ .

In the c.m. frame of the colliding hadrons, introducing the rapidity $y=\tanh^{-1}(\frac{P_z}{E})$ and
the transverse momentum $\vect P_T$, we obtain the double-differential cross section in  $P_T$ and $y$
from \ce{eq:Edsdp3}:
\eqs{
\frac{d\sigma}{dydP_T}=\int_0^1 dx_1 dx_2 g(x_1) g(x_2) 2 \hat s  P_T\frac{d\sigma}{d\hat t}
\delta(\hat s+\hat t+\hat u-M^2)&&.
}

At this stage, we can perform the integration on $x_2$ (or $x_1$) using the delta function.

In terms of the transverse energy $E_T=\sqrt{P_T^2+M^2}$, we get
\eqs{
\hat s=s x_1x_2, \ 
\hat t = M^2-x_1e^{-y}\sqrt{s}E_T,\ 
\hat u =M^2-x_2e^{y}\sqrt{s}E_T,
}
so that we obtain 
\begin{equation}\label{eq:x1x2}
x_2=\frac{x_1E_T\sqrt{s}e^{-y}-M^2}{\sqrt{s}(\sqrt{s} x_1-E_Te^{y})}.
\end{equation}
The double differential cross section on $P_T$ and $y$ then takes the following form:
\eqs{
\frac{d\sigma}{dydP_T}=\int_{x_1^{min}}^1 dx_1 \frac{2 \hat s P_T g(x_1) g(x_2(x_1))}
{\sqrt{s}(\sqrt{s} x_1-E_Te^{y})}
\frac{d\sigma}{d\hat t},
}
where $x_1^{min}$ corresponds to $x_2=1$ in Eq.~(\ref{eq:x1x2}):
\eqs{\label{eq:low_bound_x_1}
x_1^{min}= \frac{E_T\sqrt{s}e^{y}-M^2}{\sqrt{s}(E_T e^{-y}-\sqrt{s})}.
}

The last step is to relate the partonic differential cross section $\frac{d\sigma}{d\hat t}$ to
the amplitude calculated from our model. To this end, we use the well-known formula:
\begin{equation} \label{eq:sigmadt}
\frac{d\sigma^{pqrs}}{d\hat t}=\frac{1}{16\pi \hat s^2}|\overline{\cal M^{pqrs}}|^2,
\end{equation}
where $|\overline{\cal M^{pqrs}}|^2$ is the squared polarised partonic amplitude for $gg\to{\cal Q} g$, 
averaged only over colour for polarised cross sections, and where $p$, $q$, $r$ and $s$ 
are the helicities of the
four particles.

As we are concerned with polarisation only for the $\cal Q$, we sum 
over gluon polarisations and define, for $r=L,T_1,T_2$:
\eqs{
\frac{d \sigma_{r}}{d\hat t}=\sum_{p,q,s=T_1,T_2}\frac{d \sigma^{pqrs}}{d\hat t}.
}

Finally , we have the double-differential polarised cross section on $P_T$ 
and $y$ :
\eqs{
\frac{d\sigma_r}{dydP_T}=\int_{x_1^{min}}^1 dx_1 \frac{2 \hat s P_T g(x_1) g(x_2(x_1))}
{\sqrt{s}(\sqrt{s}x_1-E_T e^{y})}\frac{d\sigma_r}{d\hat t}.
}

\section{Results for $J/\psi$}
Before presenting our results, we need to choose a value for $\Lambda$ and
$m_c$.
Several studies have shown, in the context of relativistic quark
models \cite{Lambda}, 
that the scale of the vertex function is between 1.42 and 2.6 GeV,  and
that $m_c$ is between 1.42 and 1.87 GeV. We choose here a value of 
$\Lambda$ in the middle range, $\Lambda=1.8$ GeV (we shall see that small
variations do not affect our results much), and a value of $m_c$ equals to the
$D^{\pm}$ mass, $m_c$=1.87 GeV, in order to have a coherent treatment of all stable
charmonium states.

Setting $\sqrt{s}$ to 1800 GeV and considering the cross section in the 
pseudorapidity range $|\eta|<0.6$, we get the following results for $J/\psi$ production at CDF.
The first plot~(see \cf{fig:g_m187_l183_mrst}~(a)) shows our result ($\sigma_{TOT}$, $\sigma_T$ 
and $\sigma_L$) for $m=1.87$ GeV and $\Lambda=1.8$~GeV. 

These new contributions are 
compared with the usual LO CSM~\cite{CSM_hadron}. 

It must be stressed that (in the Feynman gauge)
the main contribution comes from the 3-point function (\ce{vf}).
As can be seen from \cf{fig:g_m187_l183_mrst}~(b), the term 
that restores gauge invariance (\ce{Gamma4}) contributes little: 
the square of its amplitude (see \cf{fig:illus_GI_break} (b)) is about
10 times smaller than the square of the amplitude containing 
only a 3-point vertex (see \cf{fig:illus_GI_break}~(a)). Furthermore, the interference term
between the diagrams of \cf{fig:illus_GI_break}~(a) and that of \cf{fig:illus_GI_break}~(b)
is negative, so that the effect of \ce{Gamma4} is to reduce the total amplitude squared.
 In Figures~\ref{fig:GIPA_jpsi_comp1}
, we show that the normalisation of the results using
the decay width has removed most dependence on the choice of parameters\footnote{Instead 
of a factor 100 of difference expected from $\left(\frac{N_{\Lambda=1.0}}{N_{\Lambda=2.2}}\right)^2$ 
we have less than a factor 2 at $P_T=4$ GeV and a factor 3 at $P_T=20$ GeV.}.
Interestingly, as figure 11 (b) shows, the dependence on $\Lambda$ is
negligible once values of the order of 1.4 GeV are taken.
\begin{figure}[ht]
\centerline{\mbox{\subfigure[$\sigma_{\rm TOT}$, $\sigma_T$ and $\sigma_L$]{
\includegraphics[height=6.1cm,clip=true]{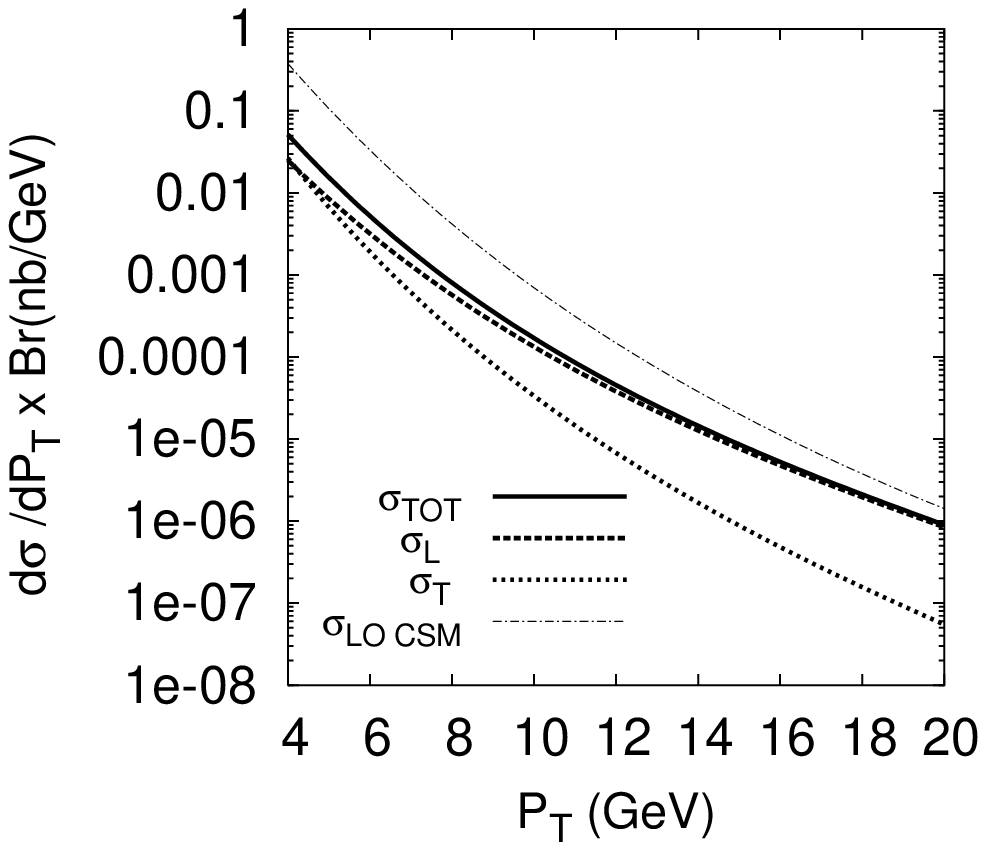}}\quad
\subfigure[$\sigma_{\rm TOT}$ and $\sigma_{\rm Pert}$,]{\includegraphics[height=6.cm,clip=true]{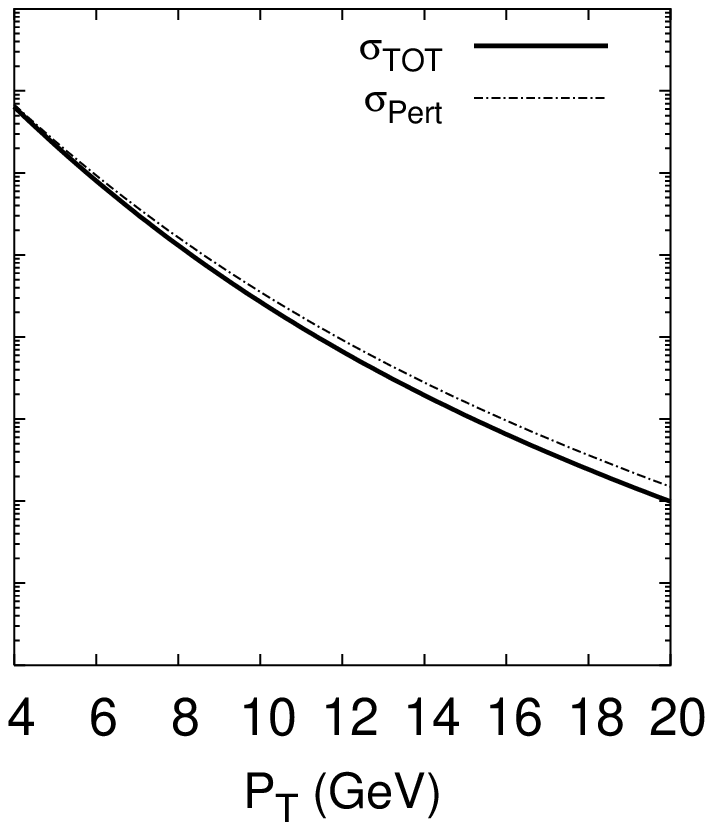}}
}}
\caption{(a) Polarised ($\sigma_T$ and $\sigma_L$) and total ($\sigma_{TOT}$) cross sections obtained 
with a Gaussian vertex function, $m=1.87$ GeV, $\Lambda=1.8$~GeV 
and the MRST gluon distribution, to be compared with LO CSM.
(b) The total (gauge invariant) contribution (plain curve) compared with that
of the 3-point vertex of diagrams of \cf{fig:illus_GI_break}~(a) (dashed curve) 
 in the Feynman gauge (CTEQ).}
\label{fig:g_m187_l183_mrst}
\end{figure} 

We see that the contribution of the new cut matters at large $P_T$.
Noteworthily, it is much flatter in $P_T$ than the LO CSM, 
and its polarisation is mostly longitudinal. 
This could have been expected as scalar products 
of $\ep_L$ with momenta in the loop 
will give an extra $\sqrt{\hat s}$ contribution, or equivalently an extra $P_T$ power in the amplitude, compared
to scalar products involving $\ep_T$. One can show that, for $\Lambda=1.8$ GeV,
$m_c=1.87$ GeV and for MRST structure functions, the longitudinal
cross section falls as $1/p_T^{8.5}$, whereas the transverse cross section
behaves asymptotically as $1/p_T^{10.5}$. The asymptotic 
power of $p_T$ changes by
5 \% for $\Lambda$ varying between 1.4 and 2.2 GeV.

Recall that in our calculation the LO CSM is zero because $M<2m$. For 
$M\geq 2m$, we should have added our 
contribution to that of the LO CSM at the amplitude level. 
The net result would then be flatter and larger.
However, it is clear that the enhancement factor would not be large enough to
reach agreement with the data.

\begin{figure}[H]
\centering
\mbox{\subfigure[Vertex function]{\includegraphics[height=4.9cm,clip=true]{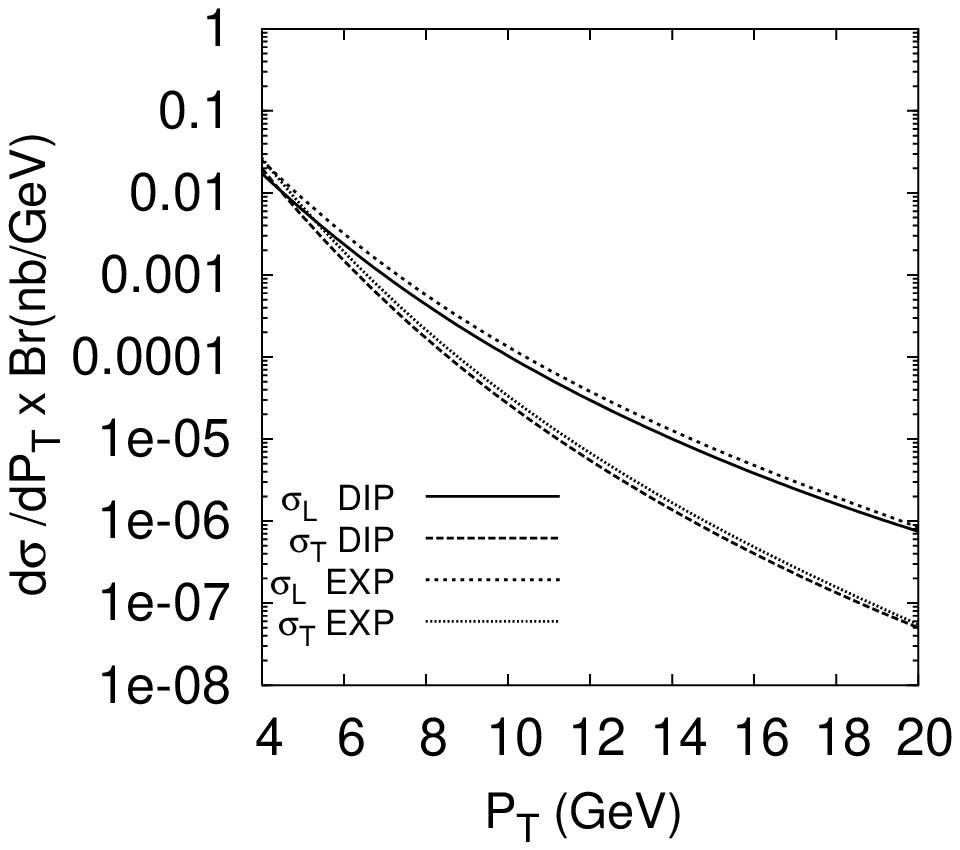}}
\subfigure[$\Lambda$]{\includegraphics[height=4.8cm,clip=true]{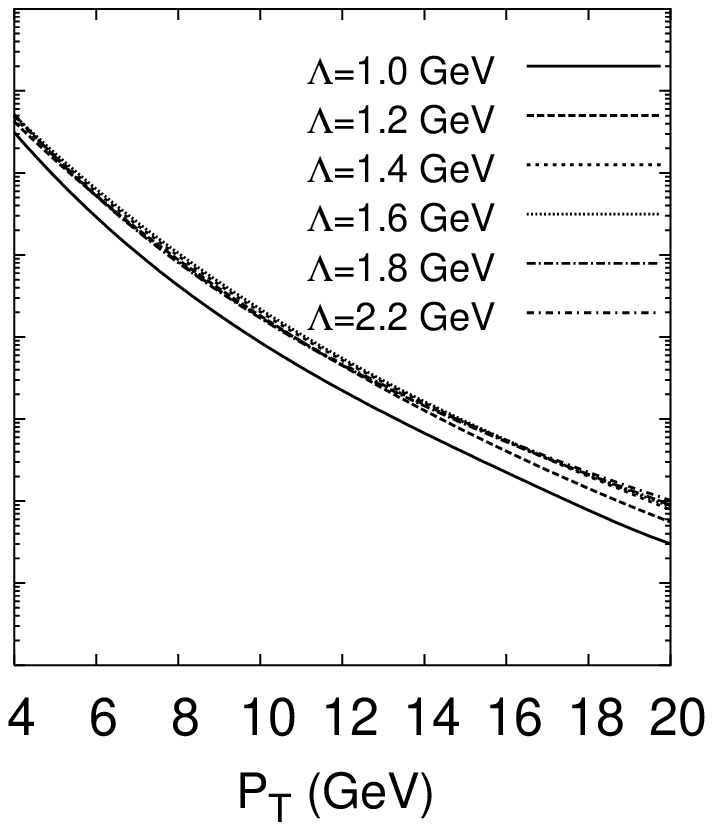}}
\subfigure[$\Lambda$ and $m_c$]{\includegraphics[height=4.8cm,clip=true]{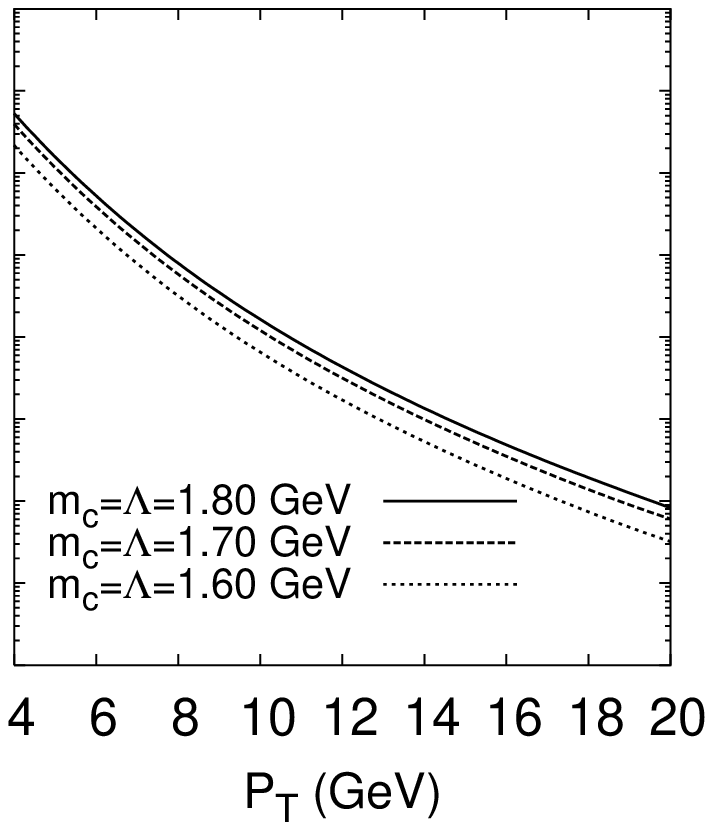}}}
\caption{(a) Comparison between the polarised cross sections obtained with the dipole and the Gaussian
vertex functions (MRST); (b) Variation of the total cross section due to a change in $\Lambda$ for a fixed value
of the quark mass (MRST); (c) Variation of the total cross section due to a change in $m$ and $\Lambda$ (CTEQ).}
\label{fig:GIPA_jpsi_comp1}
\end{figure}
\section{Results for $\psi'$}
Although the normalisation to the leptonic width 
removes most of the ambiguities in the $J/\psi$
case, it is not so for radially excited states, such as the $\psi'$.
Indeed, in this case, the vertex function must have a node. We expect it to appear
through a pre-factor, $1-\frac{|\vect p|}{a_{node}}$, multiplying the $1S$
vertex function.
Explicitly, $\Gamma_{2S}(p,P)$, for a node $a_{node}$ should be well parametrised by
\eqs{\left(1-\frac{|\vect p|}{a_{node}}\right)
\frac{N'}{(1+\frac{|\vect p|^2}{\Lambda^2})^2} \text{ or  }
N'\left(1-\frac{|\vect p|}{a_{node}}\right) e^{\frac{-|\vect p|^2}{\Lambda^2}}.}

In order to determine the node position in momentum space, we can 
fix $a_{node}$  from its known value in position space, \eg~from potential studies, and  
take the Fourier transform of the wave function. 
However, the position of the node is not very well-known, and it is unclear
how to relate our vertex with off-shell quarks to an on-shell non-relativistic
wave function. The most remarkable thing is that, because the integrand has a zero, the integral $I$ of
Eq.~(\ref{eq:I}) entering the decay width calculation can vanish for
a certain value $a_{node}=a_0$, which turns out to be close to the estimated
value of the 
zero in the wave function. Because of their different momentum dependence, the integrals that
control the production are not zero for $a_{node}=a_0$. Hence our normalisation
procedure can in principle produce an infinite answer. Of course, this means 
that one cannot be at $a_{node}=a_0$ exactly. However, if one is close to it,
then it becomes possible to produce a large normalisation. Hence in the $\psi'$
case, our procedure can produce agreement with the data at low $P_T$.

\cf{fig:g_psip_a1334_0-0} shows that for $a_{node}=1.334$ GeV, one obtains a good fit to
CDF data at moderate $P_T$ (note that the slopes are quite similar. This is at odds with 
what is commonly assumed since fragmentation processes --with a typical $1/P_T^4$ behaviour-- 
can also describe the data). The $\psi'$
is predicted to be mostly longitudinal.

\begin{figure}[H]
\centering{\includegraphics[height=6.2cm,clip=true]{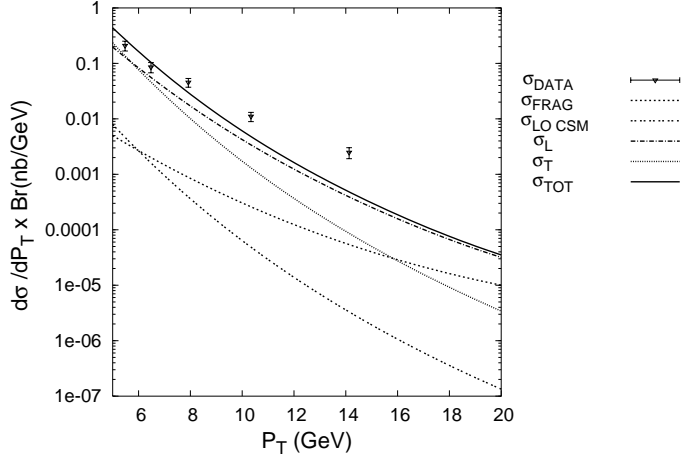}}
\caption{Polarised ($\sigma_T$ and $\sigma_L$) and total ($\sigma_{TOT}$) cross sections for $\psi'$
obtained  with a Gaussian vertex functions, $a_{node}=1.334$ GeV, $m=1.87$ GeV, \break $\Lambda=1.8$~GeV 
and the CTEQ gluon distribution, to be compared with LO CSM, CSM fragmentation~\cite{CSM_frag} 
and the data from CDF~\cite{CDF7997a}.}
\label{fig:g_psip_a1334_0-0}
\end{figure}

This effect of the node in the $\psi'$ vertex function could also solve
the $\rho-\pi$ puzzle as suggested in~\cite{Lansberg:2005ed}, since a slight modification
in the integrand of $I$ can produce a large suppression in the 
$\rho-\pi$ decay of the $\psi'$. Such a modification in the integrand
is indeed expected  to come from the presence of an off-shell $\omega$ in the $\rho-\pi$ decay
instead of an off-shell photon for the leptonic decay. On the other hand, in the case of $J/\psi$, 
no important effect are expected.

\section{Conclusion and outlook}
In this letter, we have shown that there are two singularities in the box
diagram contribution to quarkonium production. We have chosen quark masses 
so that only the $s$-channel singularity (which is usually neglected)
contributes, and vertices without cuts. 

On the theoretical side, we have begun to map the ingredients needed to go
beyond the static approximation $M=2m$. They involve the introduction of
new 4-point vertices that restore gauge invariance. We postpone a full study
of these to a later publication~\cite{article2}.

On the phenomenological side, we have shown that in the $J/\psi$ case, 
this new singularity produces  results comparable
to those of the lowest-order CSM, \ie~too small to accommodate the 
Tevatron data. 
In the $\psi'$ case, ambiguities in the position of the node of the
vertex function can lead to an enhancement, and to an agreement with the
data. Hence it is not clear that the same mechanism has to be at work
for $1S$ and $2S$ mesons.

Our approach can be used in the case of $P$ waves, where we would simply input a suitable
3-point vertex instead of taking higher derivatives of the
wave function and of the perturbative amplitude, but also in a $k_t$-factorisation framework \cite{Hagler:2000dd} 
(which would enhance the cross section in the $J/\psi$ case), and can be combined with contribution
from COM fragmentation. Because the quarkonium are mostly longitudinal in this work
and transverse in fragmentation, it seems possible to reach agreement with polarisation 
data.

\section*{Acknowledgements}
J.P.L. is an IISN Postdoctoral Researcher, Yu.L.K. was a visiting research fellow of the FNRS
while this research was conducted, and is supported by the Russian grant
RFBR 03-01-00657. We would like to thank S.~Peign\'e, M.V.~Polyakov, W.J.~Stirling and
L.~Szymanowski for useful comments and discussions.


\begin{thebibliography}{99}


\bibitem{CDF7997a}
F.~Abe {\it et al.}  [CDF Collaboration],
Phys.\ Rev.\ Lett.\  {\bf 79} (1997) 572.


\bibitem{CDF7997b}
F.~Abe {\it et al.}  [CDF Collaboration],
Phys.\ Rev.\ Lett.\  {\bf 79} (1997) 578.

\bibitem{CSM_hadron} 
C-H. Chang, 
Nucl. Phys.  B {\bf  172} (1980) 425;
R. Baier and R. R\"uckl, 
Phys. Lett.  B {\bf 102} (1981) 364; 
R. Baier and R. R\"uckl,
Z. Phys.  {\bf C 19} (1983) 251.

\bibitem{CSM_frag}
M.~Cacciari and M.~Greco,
Phys.\ Rev.\ Lett.\  {\bf 73} (1994) 1586
[arXiv:hep-ph/9405241];
E.~Braaten, M.~A.~Doncheski, S.~Fleming and M.~L.~Mangano,
Phys.\ Lett.\ B {\bf 333} (1994) 548
[arXiv:hep-ph/9405407].





\bibitem{COM}
G.~T.~Bodwin, E.~Braaten and G.~P.~Lepage,
Phys.\ Rev.\ D {\bf 51} (1995) 1125
[Erratum-ibid.\ D {\bf 55} (1997) 5853]
[arXiv:hep-ph/9407339];
P.~L.~Cho and A.~K.~Leibovich,
Phys.\ Rev.\ D {\bf 53} (1996) 150
[arXiv:hep-ph/9505329];
P.~L.~Cho and A.~K.~Leibovich,
Phys.\ Rev.\ D {\bf 53} (1996) 6203
[arXiv:hep-ph/9511315].


\bibitem{Cho:1994ih}
P.~L.~Cho and M.~B.~Wise,
Phys.\ Lett.\ B {\bf 346} (1995) 129
[arXiv:hep-ph/9411303].


\bibitem{Affolder:2000nn}
T.~Affolder {\it et al.}  [CDF Collaboration],
Phys.\ Rev.\ Lett.\  {\bf 85} (2000) 2886
[arXiv:hep-ex/0004027].

\bibitem{yr_QWG}
N.~Brambilla {\it et al.},
{\it CERN Yellow Report on ``Heavy quarkonium physics''}, CERN-2005-005
[arXiv:hep-ph/0412158].


\bibitem{Burden:1996nh}
C.~J.~Burden, L.~Qian, C.~D.~Roberts, P.~C.~Tandy and M.~J.~Thomson,
Phys.\ Rev.\ C {\bf 55} (1997) 2649
[arXiv:nucl-th/9605027].

\bibitem{Lambda}
 M.~A.~Ivanov, J.~G.~Korner and P.~Santorelli,
  Phys.\ Rev.\ D {\bf 71} (2005) 094006
  [arXiv:hep-ph/0501051],
  Phys.\ Rev.\ D {\bf 70} (2004) 014005
  [arXiv:hep-ph/0311300],
  Phys.\ Rev.\ D {\bf 63} (2001) 074010
  [arXiv:hep-ph/0007169];
 M.~A.~Nobes and R.~M.~Woloshyn,
  J.\ Phys.\ G {\bf 26} (2000) 1079
  [arXiv:hep-ph/0005056].



\bibitem{Landau_eq}
  L.~D.~Landau,
  Nucl.\ Phys.\  {\bf 13} (1959) 181;
C. Itzykson, J.B. Zuber, {\it Quantum Field Theory}, McGraw-Hill, New-York, 1980.



\bibitem{diffractive} see {\it e.g.} J.~R.~Cudell and I.~Royen,
Phys.\ Lett.\ B {\bf 397} (1997) 317
[arXiv:hep-ph/9609490];
M.~G.~Ryskin,
Z.\ Phys.\ C {\bf 57} (1993) 89;
J.~R.~Cudell,
Nucl.\ Phys.\ B {\bf 336} (1990) 1.
\bibitem{DVCS}A.~V.~Radyushkin,
Phys.\ Rev.\ D {\bf 56} (1997) 5524
[arXiv:hep-ph/9704207].




\bibitem{Drell:1971vx}
  S.~D.~Drell and T.~D.~Lee,
  Phys.\ Rev.\ D {\bf 5} (1972) 1738.


\bibitem{Lansberg:2005aw}
J.~P.~Lansberg, {\it Quarkonium Production at High-Energy Hadron Colliders}, Ph.D. 
Thesis, ULg, Li\`ege, Belgium, 2005 [arXiv:hep-ph/0507175].



\bibitem{Lansberg:2005ed}
  J.~P.~Lansberg,
  AIP Conf.\ Proc.\  {\bf 775} (2005) 11
  [arXiv:hep-ph/0507184].
 

\bibitem{barger} V.D. Barger and R.J.N. Philips, {\it Collider Physics}, 
Addison-Wesley, Menlo Park, 1987.


\bibitem{Martin:2002dr}
A.~D.~Martin, R.~G.~Roberts, W.~J.~Stirling and R.~S.~Thorne,
Phys.\ Lett.\ B {\bf 531} (2002) 216
[arXiv:hep-ph/0201127].


\bibitem{Pumplin:2002vw}
J.~Pumplin, D.~R.~Stump, J.~Huston, H.~L.~Lai, P.~Nadolsky and W.~K.~Tung,
JHEP {\bf 0207} (2002) 012
[arXiv:hep-ph/0201195].

\bibitem{article2}  J.R. Cudell, Yu.L. Kalinovsky and J.P. Lansberg,  
in preparation.

\bibitem{Hagler:2000dd}
  P.~Hagler, R.~Kirschner, A.~Schafer, L.~Szymanowski and O.~V.~Teryaev,
  Phys.\ Rev.\ Lett.\  {\bf 86} (2001) 1446
  [arXiv:hep-ph/0004263];
  P.~Hagler, R.~Kirschner, A.~Schafer, L.~Szymanowski and O.~V.~Teryaev,
  Phys.\ Rev.\ D {\bf 63} (2001) 077501
  [arXiv:hep-ph/0008316].


\end{thebibliography}
\end{document}